\documentclass[manuscript]{aastex}

\begin{document}


\title{An analytical model for the 0.33 - 7.85 micron transmission spectrum of HD189733b :
        Effect of stellar spots}

\author{A. Daassou, Z. Benkhaldoun, M. Ait Moulay Larbi and Y. Elazhari}
\affil{Laboratory of High Energy Physics and Astrophysics, Physics Department, Faculty of Science Semlalia, University of Cadi Ayyad,
    P.O.B 2390, Marrakech 40000, Morocco}
\email{ahmed.daassou51@gmail.com}

\begin{abstract}
In recent years, the transit of HD189733b is the most observed among all known transiting exoplanets. The transmission spectrum of 
this planet has been measured with several instruments from Spitzer and Hubble (HST) Space Telescopes, and from several ground-based 
telescopes. In this paper an analytical theory is used to complete this spectrum from near-ultraviolet 
to infrared (0.33 - 7.85 $\mu m$). The model suggests a new approach which take into account the quantity of light transmitted 
through the planetary atmosphere and the thermal emissions from the night side of the planet, 
to describe the transmission spectrum  of HD189733b in photometric study. 
The availability of measures in different wavelengths has allowed us 
to validate our model with more efficiency.  
We found an agreement between our  transmission spectrum model of HD189733b and data from the infrared to the UV.  The model predicts  a 
value of  $R_p /R_\star$ $= 0.1516$   at $7.3 \mu m$, which is  a low value  compared to all  observations at different wavelengths.
We interpreted this value by a fluorescence emission
from sulphur dioxide ($SO_2$).  Therefore, the likely presence of these molecules in the atmosphere of HD 189733b.  
The second objective of this paper is to study the effect of starspots on the transmission spectrum of this hot-Jupiter.  
To reach this goal, we developed an analytical theory considered as an extension of the approach proposed by \citet{Berta}.     
The model shows clearly that the unocculted spots would significantly increase 
the transit radius ratio at visible and near-ultraviolet wavelengths, while having a minimal impact at infrared wavelengths.
Therefore, the wavelength dependence of the spots effect has been clearly shown by this new model.  
At the end of this paper, we reported the way in which this model can  provide an estimation of the percentage of the unocculted 
spots area  relative to stellar disk area for an observation of HD189733 performed in a given epoch and at a given wavelength.   
\end{abstract}

\keywords{planetary systems - starspots - stars: individual: HD 189733 - planets: atmospheres - techniques: photometric }

\section{Introduction}
The increasing of the number of exoplanets discovered during last years, revealing several types of systems with différent properties. 
Our understanding of planets outside the solar system is revolutionising from various studies on the transiting planets. 
The phenomenon of transiting planets is produced when the planet is located on the same plane containing the star and the earth, during 
this phenomenon, a part of the starlight  is blocked by the planets and producing a specific light curve.    
The study of  transiting planets gives an idea for their structure. Indeed,  the transit technique can estimate the radius of the planet 
and its orbital inclination,  and coupled with data from the radial velocity technique gives an estimate of the mass. Therefore, 
the density and structure can be inferred.\

The composition of the planetary atmospheres can also be inferred from the transit technique. 
The optical depth in the atmosphere is sensitive to molecular and atomic absorption and we can say that it is dependent of wavelength. 
Therefore, the transit depth and the planetary radius  are wavelength dependent. The measure of the radius of a planet as a function 
of wavelength can allow the identification of some absorption features in the atmospheres leading  to infer the presence of molecular  
and atomic species (\citet{Brown et al.1}; \citet{Seager2}).\ 

Two approaches are used to measuring the dependence of wavelength  of a planetary radius. The first approach is by monitoring the transiting 
planets with spectroscopic measurements, which allow a measurement of transit depth at each wavelength  
by making a separation of light curves  for each channel of wavelength.
The second approach, is to perform multi-wavelength photometric 
observations of planetary transits, and estimate the transit depth for each. The model presented in this paper is focused on 
this second method.\

A new class of planets having masses identical  to Jupiter and on orbits of very short-period are called Hot Jupiters, their atmospheres 
are very hot because they are intensely irradiated by their stars.
Therefore, the large atmospheric heights of these planets make them specially good targets for the measurements  described above.
Typically, HD189733b represents  a good example for these measurements. The planet orbits a bright K2V star and characterized by a 
transit depth of $\sim2.5\%$ (\citet{F. Bouchy et al.}). Following \citet{Knutson 1,Knutson 2}  the planet has 
a brightness temperature varies between 960 and 1220 K. 
The large scale height of the atmosphere (around 200 Km) allowing transits to 
measure its chemical composition. The two approaches of measurements described above have been used to detect several species 
in the atmosphere of this hot-Jupiter, for example, sodium  detected  with ground-based observations (\citet{Redfield}).
The presence of $H_2 O$ has been inferred from Spitzer observations (\citet{Tinetti et al.}).
The detection of water and methane reported by \citet{Swain et al.1} in spectroscopic observations between 1.5 and 2.5 $\mu m$ has 
been challenged by \citet{Sing} with new HST observations at 1.66 and 1.8 $\mu m$.
The observations also suggested the existence of a haze in the upper part of the atmosphere of this hot Jupiter 
at optical wavelengths (\citet{Pont1}).\ 

Note that HD 189733 is an active star. In general, the presence of stellar magnetic variability, 
caused by cool spots, bright faculae, or magnetic surface inhomogeneities, can modify the transit
depth and have a significant impact on the derivation of the dependence of the planetary radius on the wavelength
(\citet{Pont1}; \citet{Czesla}; \citet{Sing}; \citet{Agol}; \citet{Berta}; \citet{Sing2011a}; \citet{Désert2}). 
In visible light, the presence of spots in the  photosphere of HD 189733 causes a variation  of its flux by $\pm 1.5\%$ over 
its $11.953 \pm 0.009$ days  period of rotation  (\citet{HenryandWinn}; \citet{Winn et al.}; \citet{Miller-Ricci}; \citet{Croll}).
Therefore, the existence of such spots during the transit of its planetary companion, affect the determination of the
radius of the planet and the analysis of its transmission spectrum.
Several models  have been developed by \citet{Czesla}, \citet{Ballerini}, \citet{Berta}, \citet{Désert2} to estimate the starspots  
effects on the exoplanet sizes determination.\   

In this paper, we present a new approach to study the effect of starspots on the planetary transmission spectrum.
This approach based on the representation of the stellar spectrum by the blackbody radiation described by the Planck function. 
In general, when working with photometry (and not with spectroscopy)  we observe a deviation of the stellar spectrum from 
a blackbody,  but  to model the planetary transmission spectrum by considering the stellar spectrum as a blackbody radiation, 
we need to search the analytical expression of planetary optical depth consistent with these radiation, and the combination of 
both can compensates the deviation  observed initially and  provides finally a correct expression  describing   
the planetary transmission spectrum.\\    

In the Sect. \ref{1}, we present the state of the art of the starspots effects on the radius determination of HD189733b, 
and the recent main approaches studying these effects.
The Sect. \ref{2} give a detailed description of the model for both cases : with and without  stellar activity. A comparison with 
observations will be presented in Sec. \ref{3}  in order to check the validity of our model.  
The analyze of the impact of starspots on the planetary transmission 
spectrum  is carried out also in this section. We give a conclusion in Sect. \ref{4}.     

\section{Effect of stellar spots}\label{1}
\subsection{State of the art of the starspots effects on the radius determination of HD189733b}
Since the discovery of its planetary companion (\citet{F. Bouchy et al.}), HD 189733 has been the subject of intense observations 
in both the optical bands (\citet{Bakos}; \citet{Winn et al.}; \citet{Pont}; \citet{Sing2011b}) and the infrared
wavebands (\citet{Beaulieu}; \citet{Sing}; \citet{Désert2}). It is a K-type star with strong chromospheric activity
(\citet{Wright}), which produces a quasi-periodic optical flux variation of $\sim 1.3\%$ (\citet{Winn et al.})
due to the rotation of a spotted stellar surface. The first accurate determination of the planetary system parameters was carried 
out by \citet{Bakos} through $BVRI$ photometry; neglecting the effects of stellar variability, their planet-to-star radius ratio 
($0.156 \pm 0.004$) was found to be smaller than that of \citet{F. Bouchy et al.} ($0.172 \pm 0.003$). 
\citet{Winn et al.} obtained radius measurements compatible with those of \citet{Bakos}, again neglecting the
effects of stellar variability. Even \citet{Pont} obtained results compatible with those of \citet{Bakos}, in this case
correcting their HST/ACS observations for the stellar variability; moreover, these authors expected that the starspots effect
be reduced in the infrared because of a lower spot contrast. Through HST/NICMOS transit observations, correcting for the
unocculted starspots, \citet{Sing} found a planet-to-star radius ratio of $0.15464 \pm 0.00051$ and $0.15496 \pm 0.00028$ at 1.66
and 1.87 $\mu m$, respectively, in agreement with the planet-to-star radius ratio of \citet{Bakos}.
In a multiwavelength set of HST/STIS optical transit light curves, \citet{Sing2011b} demonstrated the dependence of the
planet-to-star radius ratio on the out-of-transit stellar flux showing that the transit is deeper when the star is fainter 
as expected as a consequence of unocculted spots present on the disc of the star during the transit. 
In Spitzer/IRAC wavebands, both \citet{Beaulieu} and \citet{Désert2} detected variations in the planet-to-star radius ratio. 
The former authors found that the infrared transit depth is smaller than in the optical, and the effect of stellar spots is 
to increase this depth. By correcting for this effect, they found that the transit depth is shallower by about $0.19\%$ 
at 3.6 $\mu m$, and $0.18\%$ at 5.8 $\mu m$. In the Spitzer/IRAC's 3.6 $\mu m$ band, the latter authors observed a greater 
$R_p/R_{\star}$ ratio in low brightness periods because of the presence of starspots, implying that the apparent planet-to-star 
radius ratio varies with stellar brightness from $0.15566^{+0.00011}_{-0.00024}$ to $0.1545 \pm 0.0003$ (\citet{Désert1}) 
from low to high brightness periods, respectively.\

The above studies illustrate how stellar magnetic activity has a significant impact on the transit depth and in the 
consequent derivation of the planet radius, in both the optical and the infrared wavebands. 
\subsection{The recent main approaches studying the starspots effects on the exoplanet sizes determination}
Several approaches have been developed to study  the effect of starspots on the exoplanet sizes. A remarkable approach treating 
this effect  was carried out by \citet{Czesla} on the active star CoRoT-2.  They proposed a method for deriving the unperturbed 
transit profile by extrapolating the observed profiles towards their lower flux limit.  They pointed out also the importance of the 
normalization of the transit profile to a common reference level to make different transits comparable
with each other and show that the relative light loss during a transit is correlated significantly with the out-of-transit flux.\

\citet{Ballerini}  used a systematic approach to quantify the flux variations of the star due to its magnetic activity, as a function 
of wavelength bands. In their approach, they assumed a star with spots covering a given fraction of its disc and model the variability 
in both the $UBVRIJHK$ photometric system and the Spitzer/IRAC wavebands for dwarf stars from G to M spectral types. Thereafter, 
they compare activity-induced flux variations in different passbands with planetary transits and quantify how they affect the 
determination of the planetary radius.\ 

Another active star GJ1214 studied by \citet{Berta}.  They used a simple model 
to estimate the  transit depth variations $\Delta D_{spots}(\lambda, t)$ induced by the unocculted spots.  
They assumed a fraction $s(t)$ of the star's Earth-facing hemisphere is covered with spots, $s(t)$ will change as the star 
rotates and the spots evolve. They introduced the following definition to derive the  transit depth variations due to spots 
\begin{equation}
D(\lambda, t) = 1 - \frac{F_{i.t.}(\lambda, t)}{F_{o.o.t.}(\lambda, t)}
\end{equation}
where $F_{o.o.t.}(\lambda, t)$ and $F_{i.t.}(\lambda, t)$ are the observed out-of-transit and in-transit spectra, respectively,   
defined as 
\begin{equation}\label{oot}
 F_{o.o.t.}(\lambda, t) = [1 - s(t)] F_{\circ}(\lambda) + s(t) F_{\bullet}(\lambda) 
\end{equation}
\begin{equation}\label{it}
 F_{i.t.}(\lambda, t) = \Big[1 - s(t) - \Big(\frac{R_p}{R_{\star}}\Big)^2\Big] F_{\circ}(\lambda) + s(t) F_{\bullet}(\lambda) 
\end{equation}
where $F_{\circ}(\lambda)$ and $F_{\bullet}(\lambda)$ are respectively  the spectrum of the unspotted photosphere and that of the 
presumably cooler spotted surface. In their calculations, they neglected limb-darkening and treated each of the two components 
presented in each above expressions ((\ref{oot}) and (\ref{it})), as having uniform surface brightness.  
In addition, they used  blackbody spectra for $F_{\circ}(\lambda)$ and $F_{\bullet}(\lambda)$,
and considered  that the planet  blocks light across a spot-free transit chord.\
A limitation of this approach  is that by considering a planet with no atmosphere.  Therefore, the quantity of light transmitted 
through the planetary atmosphere has been neglected. The selective absorption of the stellar radiation by the molecular species, 
present in the planetary atmosphere, may produce a variation in the apparent radius of the planet versus the central wavelength 
of the passband.\

In this paper,  we develop the above approach proposed by \citet{Berta}.  We will keep the 
same considerations used in this approach, except that we will take into account, in this time, the quantity of light transmitted 
through the planetary atmosphere to estimate the starspots effect on the planetary transmission spectrum (see Sect.\ref{2}).    

\section{Description of the model}\label{2}   
In this study, we present a new approach describes the transmission spectrum of HD 189733b.  
Our goal  is to model this spectrum by taking into account of two 
sources of radiation emitted by the planet when it passes in front of its star : the thermal emissions from the night side of the planet 
and the quantity of light transmitted through its atmosphere.\\
The model is focused on the study of this system in a wavelength range from near-ultraviolet to infrared (0.33 - 7.85 $\mu m$), where two cases could  be taken into 
consideration. First,  when the star presents no spots on its visible surface. Second, when the flux emitted by the star is perturbed 
by the stellar activity (with starspots).           
\subsection{The wavelength dependence of the transit depth} 
In this section,  we consider that  the flux emitted by the star is  not affected by the stellar activity. 
Then, as considered by \citet{Berta}, we approximate the stellar spectrum  ($W$/$m^3$) from the Planck law of blackbody radiation : 
\begin{equation}
L_\star(\lambda)=\frac{2\pi hc^2}{\lambda^5} \frac{1}{exp\big(\frac{hc}{\lambda k_B T_\star}\big) -1}               
\end{equation} 
where $T_\star$ is the stellar effective temperature, $\lambda$ is the wavelength, 
$k_B$,  $c$ and $h$ are respectively the Boltzman constant, the light speed  and the Planck constant.\\  
In order to  integrate the contribution of the thermal emissions from the night side of the planet in the transit depth calculation, 
these emissions can be represented by :  
\begin{equation}
L_p(\lambda)=\frac{2\pi hc^2}{\lambda^5} \frac{1}{exp\big(\frac{hc}{\lambda k_B T_p}\big) -1} ~\epsilon_p               
\end{equation} 
where $T_p$ is the surface temperature of the night side of the planet,
$\epsilon_p$ the emissivity of the night side of the planetary atmosphere. In general, $\epsilon_p$ is wavelength dependent,
but as first approximation we consider it invariant in the wavelength range studied in this model (0.33 - 7.85 $\mu m$).\

From the above approximations, we define the relative transit depth as :
\begin{equation}\label{rel}
\left(\frac{\Delta F}{F}\right)_{\lambda} = \frac{F_{out} (\lambda) - F_{in} (\lambda)}{F_{out} (\lambda)}               
\end{equation} 
where $F_{in} (\lambda)$ and $F_{out} (\lambda)$  are the radiations received on the Earth per unit surface of a detector and per unit of 
wavelength ($W$/$m^3$), during the transit (from both the star and the planet) and out of the transit (from the star only) 
respectively.\ 
The maximum of information on the planetary atmosphere is obtained when the planet is in the middle of the transit, i.e.  
when its atmosphere is entirely projected on  the visible face of the star and that the limb-darkening effects of the star are minimized.\

As considered by \citet{Berta}, we neglect the limb-darkening of the star and treat it as has uniform surface brightness.
Therefore, $F_{in} (\lambda)$ and $F_{out} (\lambda)$ defined in Eq.(\ref{rel}) can be expressed as :
\begin{equation}\label{F_out}
F_{out} (\lambda) = \frac{S_\star L_\star(\lambda) }{f_\star \pi d_\star ^2} 
\end{equation} 
\begin{equation}\label{F_in}
F_{in} (\lambda) = \frac{(S_\star - S_p) L_\star(\lambda)}{f_\star \pi d_\star ^2} + \frac{S_p L_p(\lambda)}{f_p \pi d_p ^2} + 
\frac{S_p^{atm} F_T (\lambda) }{f_p \pi d_p ^2}
\end{equation} 
where $S_\star$ and $S_p$ are respectively the stellar and planetary disk areas.
$S_p^{atm}$ is the effective area allowing the transmission of light through the limb of the planet during a primary transit. 
$F_T (\lambda)$ is the quantity of light transmitted by the effective area  
$S_p^{atm}$,  $d_\star$ and $d_p$ are respectively  the Earth-Star and Earth-Planet distances  ($d_\star$ $\sim$ $d_p$).  
$f$ is a factor which depends of the anisotropy of the radiation.  
For 
the star, we consider that the radiations are emitted isotropically from its surface. 
But for the planet, it should be noted that  its atmosphere is not calm. However, the strong stellar gravity makes one hemisphere of the 
planet constantly faces the star, heating permanently only on one side. This probably creates fierce winds sweeping from 
the day side to the night side (\citet{Knutson 2}). 
Therefore, the quantity of light traversing through the 
planetary atmosphere during the transit and the thermal emissions from the night side of the planet, may  affected  by  these fierce winds 
which may cause a deviation of a part of these radiations, and leading to a non-isotropic emission from the planet.\\

The stellar and planetary surfaces appearing in Eq.(\ref{F_out}) and Eq.(\ref{F_in}), will be presented as a ratio 
$S_p/S_\star$ in the relative transit depth expression. Considering a circular orbit for the planet, this ratio can be expressed in terms of 
some parameters derived directly from the transit light curve using the following analytic approximations (\citet{Carter}, 
\citet{Seager1}):  
\begin{equation}\label{b}
 b^2 \approx 1 - \frac{T}{t_0} \times \frac{R_p}{R_\star}, 
\end{equation}
\begin{equation}\label{semi-axe}
 \frac{a}{R_\star} = \frac{b}{\cos i}  \approx \frac{P\sqrt{\frac{R_p}{R_\star}}}{\pi \sqrt{Tt_0}},
\end{equation}      
where $t_0$ is the egress or ingress  duration ($t_0 = t_{II}$ - $t_{I} $),  $t_{II}$ and $t_{I}$ are the time of second and first 
contact, $T$ is the total transit duration,  $R_\star$ and $R_p$ are the stellar and planetary radii, $a$ is the orbital semi-major 
axis, $i$ is the orbital inclination, $P$ is the orbital period and $b$ is the impact parameter.\

From the Eq.(\ref{b}) and Eq.(\ref{semi-axe}), the planet-to-star surface ratio $S_p/S_\star$ is estimated by
\begin{eqnarray}\label{p_star}
\frac{S_p}{S_\star} \approx \left(\frac{P ^2 \cos^2 i}{\pi ^2 T t_0} + \frac{T}{t_0}\right)^{-2} 
\end{eqnarray}

The relative transit depth can thus deduced by combining the expressions (\ref{F_out}) and (\ref{F_in}) with the Eq.(\ref{p_star}) :   
\begin{eqnarray}\label{coco}
\left(\frac{\Delta F}{F}\right)_{\lambda} = \frac{S_p}{S_\star}
\Big(1 - \epsilon_p \frac{f_\star d_\star ^2}{f_p d_p ^2} \frac{(exp(\frac{hc}{\lambda k_B T_\star}\big) -1)}{(exp(\frac{hc}
{\lambda k_B T_p}\big) -1)}\Big)   -
\frac{S_p^{atm}}{S_\star} \frac{f_\star d_\star ^2}{f_p d_p ^2} \frac{F_T (\lambda)}{L_\star(\lambda)}   
\end{eqnarray}
where  $S_p/S_\star$ is estimated from the Eq.(\ref{p_star}). For gaseous giant planets, the ratio between the area which can transmit light  
 and area star-minus-planet is very low, of the order of $10^{-4}$ to $10^{-3}$ (\citet{Batista}).  This allows us to estimate the  
$S_p^{atm}/S_\star$ ratio as
\begin{eqnarray}\label{atmosph}
 \frac{S_p^{atm}}{S_\star} \approx 10^{-3} \Big(1 - \frac{S_p}{S_\star}\Big)
\end{eqnarray}

The quantity of light transmitted through the limb of the planet ($W$/$m^3$) is defined as 
 
\begin{eqnarray}\label{portion}
F_T (\lambda) =  e^{- \tau (\lambda)} L_\star(\lambda)
\end{eqnarray} 
where $\tau (\lambda)$  is the optical depth of the planetary atmosphere as a function of the wavelength. We know that the way exoplanet 
atmospheres transmit light depends (a) on their chemical composition and (b) the temperature dependent way in which the species that make 
up atmosphere absorb and emit light.  But when working with photometric models and considering the incident stellar flux as a blackbody 
radiation, we need in this case to search the transmission mechanism of this radiation and the corresponding optical depth expression 
which allow us to explain the transmission spectrum  of HD189733b.\

In this work, we consider that the transmission of light through the planetary atmosphere is done by the intermediate of  the fluorescence
phenomenon. With this in mind, we suggest that the temperature dependent way in which the species that make up atmosphere absorb and 
emit light should be represented  by a power law of the quantity $\frac{k_B T_{pl}}{h \nu}$  when the stellar spectrum is 
considered as a blackbody radiation, where $T_{pl}$ is the equilibrium temperature of the planetary atmosphere reached after  a certain 
number of absorption and emission of photons each has an energy $h\nu$.\

From the above considerations, the quantity $e^{- \tau (\lambda)}$ ,which represents the portion of the transmitted light, 
can be expressed as :             
\begin{eqnarray}\label{valis1}
e^{- \tau (\lambda)} = \sum_{k=0} ^{n} f_k \Big(\frac{k_B T_{pl}}{h \nu}\Big)^k
\end{eqnarray} 

where $n$ is a integer number ($n \in N^{\star}$), $f_k$ are proportionality coefficients correspond to different  values of $k$.
From the above expression, the optical depth $\tau(\lambda)$ can be written as :        
\begin{eqnarray}\label{valis2}
 \tau (\lambda) = -log \Big[ \sum_{k=0} ^{n} f_k \Big(\frac{k_B T_{pl}}{h \nu}\Big)^k \Big]
\end{eqnarray}
Both expressions (\ref{valis1}) and (\ref{valis2}) are valid mathematically only when the quantity 
$\sum_{k=0} ^{n} f_k \Big(\frac{k_B T_{pl}}{h \nu}\Big)^k$  is strictly positive. To examine the sign of this latter,
we suggest the following polynomial form of variable k for the proportionality coefficients $f_k$,   
which, at the same time, must verifies the mathematical condition and has a physical signification
\begin{eqnarray}
f_k = (-10)^k \sum_{i=0} ^{5} \beta_i k^i
\end{eqnarray} 
with this expression of $f_k$, the quantity  $\sum_{k=0} ^{n} f_k \Big(\frac{k_B T_{pl}}{h \nu}\Big)^k$ 
satisfied to both above conditions, in the wavelength range considered in this model (0.33 - 7.85 $\mu m$), for $k \leq 5$ ($n = 5$)
and for the following values of the coefficients $\beta_i$ : $\beta_0 = 2.053$, $\beta_1 = -15.7469$, $\beta_2 = 26.8868$, 
$\beta_3 = -13.4712$, $\beta_4 = 2.68125$, $\beta_5 = -0.18793$.

\subsection{Impact of starspots on the transit depth}\label{Impact of starspots}
   
\begin{figure}
  \centering
  \includegraphics[scale=0.7]{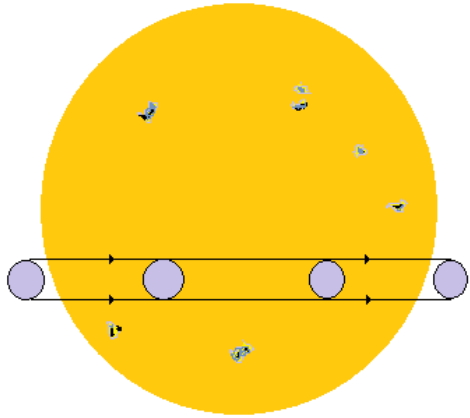}
  \caption{Geometrical view of the HD189733 system considered in this model to study the effect  of stellar spots on the transmission spectrum 
of HD 189733b.}
\label{spot1}
\end{figure}

In this section, we present a development of the approach proposed by \citet{Berta} (and by \citet{Ballerini} that they have 
used almost the same concept) to estimate the starspots effect on the planetary radius determination.
We will take into account, in this estimation, the quantity of light transmitted 
through the planetary atmosphere (described in the last section) neglected by \citet{Berta}.
As considered by these authors, we study the effect of  starspots unocculted by 
the planet which  blocks light across a spot-free transit chord (Fig. \ref{spot1}).
In addition, we neglect the effect of brighter 
active regions, faculae and plages. This effect could be important, but we note that none of the space 
and ground-based monitoring of the HD 189733b system available shows a significant signature of the crossing of a brighter region.   
  
Previous HST ACS observations of HD 189733 have established that the effective temperature of starspots is approximately 1000 K
cooler than that of the stellar photosphere (\citet{Pont1}) with a spot size 
of $\sim 2.8\%$ of the stellar disk area. 
According to the approach proposed by \citet{Berta}, we will take into account in our expressions that the spot size parameter 
is expected to evolve in time by 
introducing the parameter  $\alpha$ which represents the percentage of the spots area  relative to stellar disk area. 
The starspots have a different temperature, so to estimate their effect on the relative transit depth, we consider an 
average temperature $T_{spots}$ corresponds to each of these spots, and we treat them as a blackbody with this temperature.\

From the above considerations, the expressions of $F_{out} (\lambda)$ and $F_{in} (\lambda)$ 
become :
\begin{equation}\label{F_out ^(spots)}
F_{out} ^{spots} (\lambda) = (1 - \alpha) F_{out} (\lambda) +  \frac{\alpha S_\star L_{spots} (\lambda)}{f_s \pi d_\star ^2} 
\end{equation}  
\begin{equation}\label{F_in ^(spots)}
F_{in} ^{spots} (\lambda) = F_{in} (\lambda) - \alpha F_{out} (\lambda) + \frac{\alpha S_\star L_{spots} (\lambda)}{f_s \pi d_\star ^2} 
\end{equation}   
where $f_s$ is a factor depending on the anisotropy of the radiation emitted from the spots. 
This coefficient is assumed identical to $f_\star$ (the light is emitted isotropically from the spots).
It is important to note that the present description neglects the limb darkening of the spots.\

From the expressions (\ref{F_out ^(spots)}) and (\ref{F_in ^(spots)}), the relative transit depth defined by the Eq.(\ref{rel})   
is given by :
\begin{eqnarray}\label{spectre_spots}
\left(\frac{\Delta F}{F}\right)_{\lambda} ^{spots} &=& \frac{S_p}{S_\star}
\left[\frac{\frac{1}{exp\big(\frac{hc}{\lambda k_B T_\star}\big) -1} - \epsilon_p \frac{f_\star d_\star ^2}{f_p d_p ^2} \frac{1}{exp\big(\frac{hc}
{\lambda k_B T_p} \big) -1} }{\frac{1 - \alpha}{exp\big(\frac{hc}{\lambda k_B T_\star}\big) -1} + 
\frac{\alpha}{exp\big(\frac{hc}{\lambda k_B T_{spots}}\big) -1}} \right] -
\frac{\frac{S_p^{atm}}{S_\star} \frac{f_\star d_\star ^2}{f_p d_p ^2} \frac{F_T (\lambda)}{L_\star(\lambda)}}{(1 - \alpha) + \alpha 
\frac{exp\big(\frac{hc}{\lambda k_B T_\star}\big) -1}{exp\big(\frac{hc}{\lambda k_B T_{spots}}\big) -1}} 
\end{eqnarray}
where $T_{spots}$ represents the effective temperature of the spots. We kept respectively for $S_p/S_\star$ and $S_p^{atm}/S_\star$ 
the expressions (\ref{p_star}) and (\ref{atmosph}). Since the total transit duration ($T$) and the ingress/egress durations ($t_0$) are not 
affected by the unocculted  starspots. On the other hand, \citet{Czesla} found that no significant correction in the inclination $i$ 
due to starspots effect.  
Note that for $\alpha = 0$, we find the expression of the relative transit depth given by the Eq.(\ref{coco})
when the star presents no spots on its visible surface. 
\section{Results and discussion}\label{3} 
\subsection{Model validation}\label{Modelvalidation}
In the framework of the approximations used in this model, the wavelength dependence of the planet-to-star radius ratio can be estimated as the square root of the 
relative transit depth defined in section 2 : 
\begin{eqnarray}\label{}\nonumber
\left(\frac{R_p}{R_\star}\right)_{\lambda} &=& \left[\frac{S_p}{S_\star}
\Big(1 - \epsilon_p \frac{f_\star d_\star ^2}{f_p d_p ^2} \frac{(exp(\frac{hc}{\lambda k_B T_\star}\big) -1)}{(exp(\frac{hc}
{\lambda k_B T_p}\big) -1)}\Big)   -
\frac{S_p^{atm}}{S_\star} \frac{f_\star d_\star ^2}{f_p d_p ^2} e^{- \tau (\lambda)} \right]^{1/2} 
\end{eqnarray} 
The comparison with observations is necessary to validate our model. 
The ideal  is to compare  with simultaneous data, since the activity of the star allows  the evolution in time of  the 
spots size and their distribution on the stellar disk surface, leading to different impacts on the measure of the planetary radius. 
As we  have no simultaneous data throughout the wavelength range considered in this work (0.33 - 7.85 $\mu m$), 
the comparison will be done with the planetary transmission spectrum  measured by several instruments at different epochs 
from Spitzer and Hubble (HST) Space Telescopes, and corrected from starspots by a consistent treatment (\citet{Pont2}).\ 
  
\begin{figure*}
\begin{center}
\includegraphics[angle=0,width=6.5 in,height=4 in]{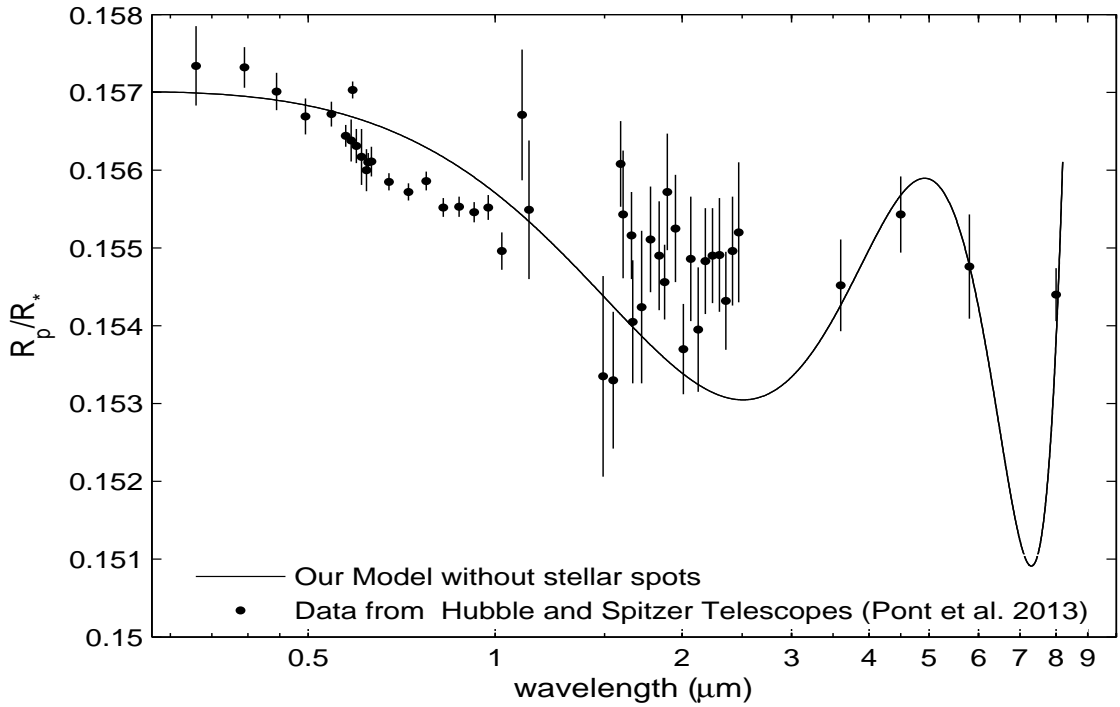}
\caption{Our transmission spectrum model of HD 189733b shown as the black curve and over-plotted by data 
from a new tabulation (\citet{Pont2}) of the transmission spectrum across the entire visible, near-ultraviolet and infrared range.   
These data have been corrected from spots effect (see section \ref{Modelvalidation})} 
\label{Model_curve}
\end{center}
\end{figure*}

In Figure. \ref{Model_curve}, 
we plot our transmission spectrum model of HD 189733b using the values presented in Table.1 correspond ​to its different  
parameters. In the same figure, the model was over-plotted by data  
from a new tabulation of the planetary transmission spectrum across the entire visible, near-ultraviolet and infrared range provided by  
\citet{Pont2}. In that table, the radius ratio in each wavelength band was re-derived, and a special care was taken to correct for, 
and derive realistic estimates of the uncertainties due to, both occulted and unocculted star spots. Details of the analysis of the data 
(and for the derivation of new error bars) can be found in \citet{Pont2}.
In general we found a remarkable agreement between our model and the data. In the mid-infrared range, we have a total agreement with 
observations  at $3.6$ $\mu m$, $4.5$ $\mu m$, $5.8$ $\mu m$ and  $7.85 $ $\mu m$. Similarly for the observations in the near-ultraviolet 
range ($0.33$ $\mu m$, $0.395$ $\mu m$, $0.445$ $\mu m$ and  $0.495$ $\mu m$). For the $0.625 - 0.925$ $\mu m$ wavelength range, the model
is on average over-predict the transit radius ratio by $0.25\%$ compared to data. However, it is on average under-predict the transit 
radius ratio by $0.84\%$ compared to observations in the $1.550 - 2.468$ $\mu m$ wavelength range.\

A remarkable prediction shown by this model at $7.3$ $\mu m$, where the  $R_p /R_\star$ ratio  has a low value  compared to all 
observations at different wavelengths presented in Figure. \ref{Model_curve}. The  predicted value is $R_p /R_\star$ $= 0.1516$.
Note that no space or ground-based observations were  acquired at this precise wavelength to test this prediction. 
Future observations of HD 189733 shall help verify this prediction. The interpretation of this low value of radius ratio at $7.3$ $\mu m$ 
should take into account the fluorescence process considered in this model. The molecules of the planetary atmosphere which can absorb and 
emit light at this wavelength will increase the total flux detected during a primary transit, which leads to have a decrease of the 
transit depth. Therefore, the radius ratio will decrease also. Among the gas-phase molecules, a strong possible candidate is emission 
from the $SO_2$  $\nu_3$ band at $7.3$ $\mu m$. \citet{Crovisier} shows that the $\nu_3$ band of $SO_2$ at 
$7.3$ $\mu m$ has a fluorescence emission rate of 6.6 $\times$ $10^{-4}$ $s^{-1}$
at 1 AU from the Sun. An absorption feature of the $\nu_3$ vibrational band of gas-phase $SO_2$ has been detected 
in the mid-infrared spectral region around $7.3$ $\mu m$ from a sample of deeply embedded massive protostars (\citet{Keane}).
In addition, the $SO_2$ excitation temperature ranging up 700 K (\citet{Keane}), which is in agreement with the equilibrium 
temperature ($T_{pl}$) considered in this model (see Table.1).
A sufficient abundance of these molecules in the atmosphere of HD 189733b, can explain the low value of the radius ratio predicted 
by this model.

\subsection{Impact of starspots on the transmission spectrum of HD 189733b} 

\begin{figure*}
\begin{center}
\includegraphics[angle=0,width=6.5 in,height=4 in]{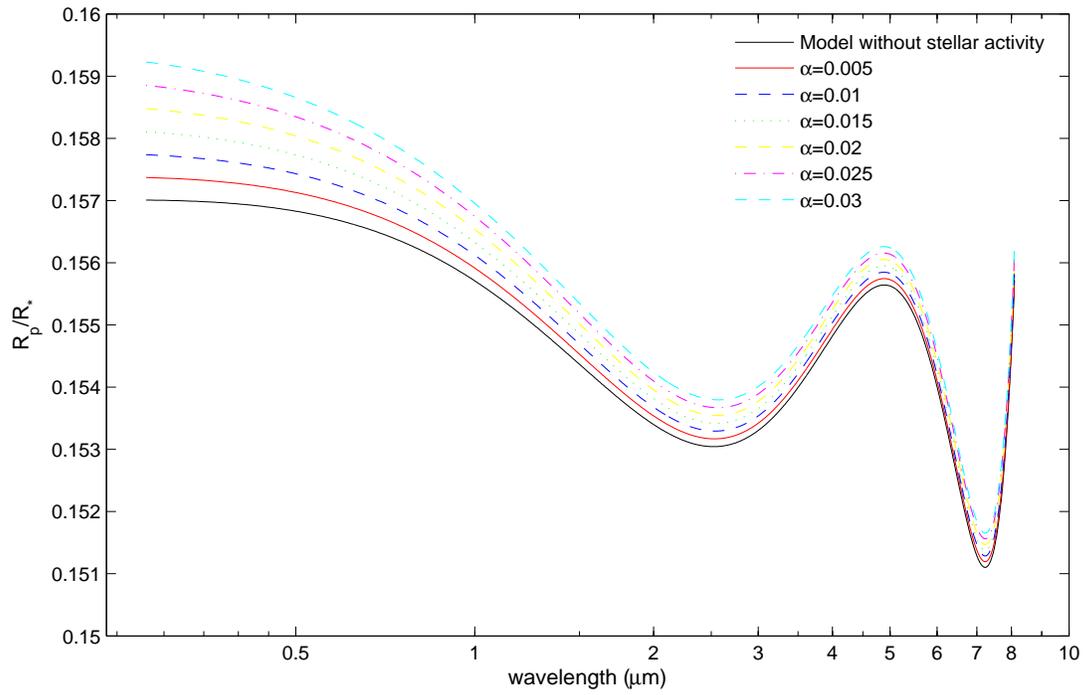}
\caption{Transmission spectrum of HD 189733b affected by the unocculted spots  taking into account of different values ​​of  $\alpha$  
from $0.5 \%$ to $3 \%$ with a step of $0.5 \%$.} 
\label{Model_spots}
\end{center}
\end{figure*}

The transmission spectrum of HD 189733b derived from the 
Eq.(\ref{spectre_spots}) is strongly depends of the percentage, represented by $\alpha$, of unocculted spots area  relative to stellar 
disk area. Several independent approaches indicate that the background spot level is between zero and $3 \%$, with lower values being 
more likely. \
In Figure.\ref{Model_spots},  we plot the transmission spectrum of HD 189733b affected by the unocculted spots for different values ​​of  
$\alpha$  from  $0.5 \%$ to $3 \%$ with a step of $0.5 \%$. In order to analyze the evolution of this transmission spectrum as a function 
of the wavelength and of $\alpha$, we take for example the two extreme values of $\alpha$ considered in this paper. 
For $\alpha$ = $0.5 \%$,  and by comparing with our model without spots, the planet-to-star radius ratio is overestimated 
by $0.20 \%$  at $0.33$ $\mu m$, by $0.13 \%$  at $0.6$  $\mu m$ and by $0.064 \%$ at $4.85$ $\mu m$.
Whereas for  $\alpha$ = $3 \%$,  the radius ratio  is overestimated by $1.34 \%$, $1.02 \%$ and $0.45 \%$ at   $0.33$ $\mu m$  , $0.6$ $\mu m$ 
and $4.85$ $\mu m$  respectively. The same figure shows that  we have a lower impact on the transmission spectrum at the 5.5 - 8 $\mu m$  
wavelength range and this for all values ​​of $\alpha$ considered in this work.\
  
From these results, it is clear that for a given wavelength, the planet-to-star radius ratio increases with $\alpha$.  Therefore, 
the effect on the transmission spectrum of HD189733b becomes  important when the star presents more spots on its visible surface. 
In addition, figure.\ref{Model_spots} shows also that this effect is wavelength-dependent,  since the unocculted spots would significantly increase 
the transit radius ratio at visible and near-ultraviolet wavelengths, while having a minimal impact at infrared wavelengths.\\
This new analytical model was able to explain well the influence of unocculted starspots on the transmission spectrum of HD 189733b 
in the UV-to-IR wavelength range. One of the significant results showed by this model is that we have a negligible impact of 
starspots on the transmission spectrum of this planet for any observation made at wavelengths tending to 8 $\mu m$.\

The model can also estimate the percentage of the unocculted spots area  relative to stellar disk area ($\alpha$).  For an observation 
performed in a given epoch with a given instrument and at a given wavelength, this percentage can be deduced by knowing the value of the 
difference between the transit radius ratio uncorrected for the unocculted spots ($(R_p/R_s)_{uncorr}$) derived from the observation 
and its corresponding value 
corrected for spots ($(R_p/R_s)_{corr}$) from the data itself which can provide a constraint of the spots effect, independently of 
the models of spots correction. Thus, the same value of this difference, but this time found by making the difference between 
the transit radius ratio affected and not affected by spots derived from our model, can estimate the value of $\alpha$.\

\section{Conclusions}\label{4} 
An analytical model has been presented in this paper to explain the transmission 
spectrum of HD189733b from UV to IR (0.33 - 7.85 $\mu m$). We found a remarkable agreement between the model and the data.     
The model predicts  a 
value of  $R_p /R_\star$ $= 0.1516$   at $7.3 \mu m$, which is  a low value  compared to all  observations at different wavelengths.
We interpreted this value of the radius ratio by a fluorescence emission
from sulphur dioxide ($SO_2$). Therefore, the likely presence of these molecules in the atmosphere of HD 189733b.
    
The current model represents an extension of the approach proposed by \citet{Berta} 
to study the effect  of stellar spots on the planetary transmission spectrum,
and by using it we found that the unocculted spots have a remarkable influence on the transit radius ratio at ultraviolet and visible 
wavelengths, while having a minimal impact at infrared wavelengths. Therefore, the wavelength dependence of this effect is clearly showed 
 by our analytical model. This model can also provide an estimation of the percentage of the unocculted spots area  
relative to stellar disk area for an observation of HD189733 performed in a given epoch and at a given wavelength.\ 

\section*{Acknowledgments}
The authors wish to thank Frederic Pont for  the data  used in this work  to validate our model.  Likewise, our sincere gratitude  
to all LPHEA team for their help and sympathy.

\clearpage
\begin{table*}
\center
   \begin{tabular}{c c c c}
\hline 
   
    & Parameter  & Value & Reference
  \\ \hline \\
    
    & $i$           & $85.61 \pm 0.04$ degrees& \citet{Knutson 2}\\
    & $P$           & $2.2$ days              & \citet{Knutson 2}\\ 
    & $T$           & $1.8$ h                 & \citet{Knutson 2}\\ 
    & $T_\star$     & $5000 K$                & \citet{Pont2}\\ 
    & $T_{spots}$   & $4000 K$                & \citet{Pont2}\\  
    & $T_p$ = $T_{pl}$      & $1000 K$        & Approximation used in this model\\             
    & $f_\star/f_p$ & $1/2$                   & inspired from our paper \citet{Bouley}\\       
    & $\epsilon_p$  & $1/2$                   & hypothesis used in this model\\
    \\ \hline
     \end{tabular}
      \label{alpha}
      \vspace{0.3cm} \caption{The values ​​of different parameters used to plot the model}
      \end{table*}
\clearpage
 
\clearpage 

\clearpage

\clearpage

\end{document}